# Associations between iron and mean kurtosis in iron-rich grey matter nuclei in aging

**Jason Langley[1+], Kitzia Solis[2], Vala Masjedizadeh[3], Murphy Shao[1], Ilana Bennett[2], and Xiaoping P. Hu[1,3]**

[1] Center for Advanced Neuroimaging, University of California Riverside, Riverside, CA, USA
[2] Department of Psychology, University of California Riverside, Riverside, CA, USA
[3] Department of Bioengineering, University of California Riverside, Riverside, CA, USA

+jason 'dot' langley 'at' ucr.edu

## Abstract

Elevated kurtosis values have been observed in subcortical grey matter structures of patients with neurodegenerative diseases. Here we tested whether these elevated values are related to iron content, which generate magnetic fields that add to the diffusion encoding gradients. Multi-shell diffusion and multi-echo gradient echo acquisitions were used to derive mean kurtosis and iron measures ($R_2$* and magnetic susceptibility), respectively, in subcortical grey matter nuclei and white matter tracts in a discovery cohort (110 older and 63 younger adults) and replication cohort (72 healthy older adults). Iron-rich grey matter regions exhibited higher mean kurtosis, $R_2$*, and magnetic susceptibility and white matter regions had lower mean kurtosis in the older adult group. In both cohorts, mean kurtosis was significantly correlated with $R_2$* and magnetic susceptibility in iron-rich grey matter nuclei. No association was seen between signal-to-noise ratio and mean kurtosis in any grey matter region, indicating that the increase in mean kurtosis was not due to reduced signal-to-noise. Finally, a phantom experiment found higher mean kurtosis as iron concentrations increased. Our findings indicate that kurtosis is associated with iron-sensitive metrics in iron-rich grey matter structures, suggesting that kurtosis may be sensitive to iron deposits.

## 1. Introduction

Diffusion-weighted magnetic resonance imaging (MRI) techniques have been used to study effects of normal aging[1-3] or disease[4-6] on brain tissue microstructure. Diffusion tensor imaging (DTI) is a commonly used approach that assumptions diffusion is Gaussian. However, the cells, organelles, and cell membranes that comprise tissue can impede diffusion and cause it to be a non-Gaussian process. Diffusion kurtosis imaging (DKI) is able to characterize non-Gaussian diffusion by relaxing this assumption[7] and measures the degree to which diffusion differs from a Gaussian process, known as kurtosis. Higher mean kurtosis (MK) values are thought to be indicative of more complex tissue microstructure[8,9].

Application of DKI to aging has consistently revealed lower kurtosis values in white matter tracts of older adults relative to younger adults[10-15], consistent with evidence that aging is accompanied by degradation of white matter[16,17], which would remove barriers to diffusion and manifest as lower MK[8,18]. In contrast, conflicting results have been seen in subcortical grey matter nuclei, with lower MK in older than younger adults in the caudate and thalamus, but higher MK in older than younger adults in the putamen[11,12]. Higher MK has also been seen in the globus pallidus, putamen, and substantia nigra in participants with neurodegenerative disorders (Parkinson's Disease) relative to age-matched controls[19,20]. Whereas the mechanism for this increase is unknown[19], its occurrence in grey matter nuclei that accumulate iron throughout the lifespan[21-24] suggests that iron content is related to kurtosis. Iron-sensitive MRI metrics ($R_2$*, magnetic susceptibility) correlate with DTI metrics (mean diffusivity, fractional anisotropy) within iron-rich grey matter nuclei in aging and disease[25-27]. The relationship may be related to an interaction between diffusion encoding gradients and the magnetic fields generated by iron deposits[28,29] or it may be due to physiological changes associated with iron deposition. However, the effect of iron on diffusion kurtosis remains unknown.

The purpose of the present study was to characterize the relationship between iron deposition and mean kurtosis values in subcortical grey matter. In a discovery cohort of younger and older adults, we assessed age group differences in iron-sensitive MRI metrics and MK in subcortical grey matter structures as well as in white matter tracts and then examined relationships between measures of iron and MK in subcortical grey matter structures. Relationships between iron-sensitive MRI metrics and MK were then repeated in a replication cohort of healthy older adults from the Alzheimer's Disease Neuroimaging Initiative database. Finally, to further test this relationship, we measured kurtosis in an agarose phantom with four levels of iron concentrations.

## 2. Methods

### *2.1 Discovery Cohort*

*2.2.1 Participants.* A cohort consisting of 110 older adults (65 female/45 male; mean age=69.3 years ± 6.3 years; age range: 60 years–87 years) and 63 younger adults (42 female/21 male; mean age=20.2 years ± 2.0 years; age range: 18 years–28 years) participated in this study. Each participant gave written informed consent prior to enrolling in the study as approved by the local institutional review board.

*2.1.2 Structural acquisition and processing.* Images from a $T_1$-weighted MP-RAGE sequence (echo time (TE)/repetition time (TR)/inversion time=3.02/2600/800 ms, flip angle=8°, voxel size=0.8×0.8×0.8 mm$^3$) were used for registration from subject space to common space. $T_1$-weighted images were analyzed with FMRIB Software Library (FSL). A transformation was derived between individual subject space to MNI152 $T_1$-weighted space using FMRIB's Linear Image Registration Tool (FLIRT) and FMRIB's Nonlinear Image Registration Tool (FNIRT) in the FSL software package[30,31].

*2.1.3 Diffusion acquisition and processing.*

Diffusion-weighted data were acquired using a spin-echo EPI with twice-refocused (bipolar) diffusion encoding gradients using the following parameters: TE/TR = 102/3500 ms, FOV = 212×182 mm, voxel size = 1.7×1.7×1.7 mm, 64 axial slices, six $b$=0 images, multiband acceleration factor = 4, and adaptive combine was used to combine signals from each channel in the 32-channel head coil. Diffusion-weighting was applied in 64 directions per $b$-value with two $b$-values ($b$=1500 s/mm$^2$ and $b$=3000 s/mm$^2$). Six b=0 images with reversed phase-encoding were acquired to correct for susceptibility distortions.

Diffusion data were denoised using dwidenoise in mrtrix [32], then corrected for motion, eddy current and susceptibility distortions using eddy in FSL [33]. Next, skull stripping of the $T_1$-weighted image and susceptibility corrected b=0 image was performed using the brain extraction tool in the FSL software package [34]. Mean diffusivity (MD) was estimated using dtifit in FSL and MK was estimated in DIPY [35]. The brain mask was dilated and the area outside the dilated brain mask was used to calculate the noise term in the signal-to-noise ratio (SNR) calculation for the b=3000s/mm$^2$ shell using the noise estimate from dwidenoise. Signal from the skull and extracranial regions were excluded from SNR estimation and each mask and signal histogram was inspected to confirm that the ROI did not include any high signal regions. Denoised images are shown in **Figure 1**.

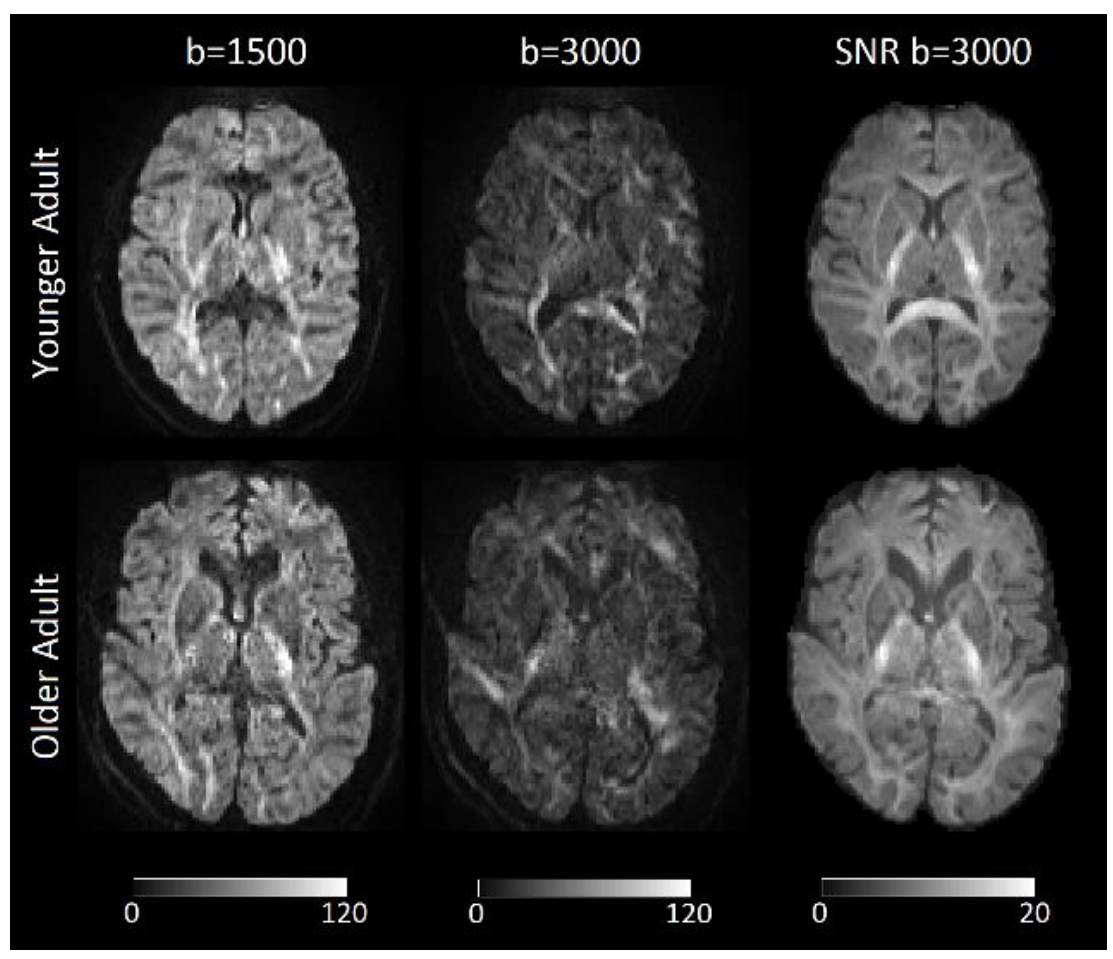

**Figure 1.** Denoised images from the $b$=1500 s/mm$^2$ (left column) and $b$=3000 s/mm$^2$ (middle column), and $b$=3000 s/mm$^2$ SNR images (right column) in a younger adult (top row) and older adult (bottom row) participant.

*2.1.4 Iron acquisition and processing.* Multi-echo data were collected with a 6-echo 3D gradient recalled echo sequence: TE1/ΔTE/TR =4/6/40 ms, FOV=192×224 mm$^2$, matrix size=192×224×96, and slice thickness=1.7 mm. $R_2$* was calculated from the magnitude data assuming a mono-exponential decay.

*2.1.5 Regions of interest.* Globus pallidus, putamen, caudate nucleus, thalamus, and hippocampus regions of interest (ROIs) were defined using the Harvard-Oxford subcortical atlas and the dentate nucleus was defined using a previously published atlas[36]. MK and mean $R_2$* were measured in each grey matter ROI. White matter ROIs for the superior longitudinal fasciculus, forceps major, and forceps minor were defined using the Johns-Hopkins white matter atlas and mean MK was measured in each ROI. For each ROI, voxels with MK values that were negative or greater than 4 were excluded.

### *2.2 ADNI Replication Cohort*

*2.2.1 Database and participants.* Data from the Alzheimer's Disease Neuroimaging Initiative (ADNI) database (adni.loni.usc.edu) were used to replicate the observed relationship between iron and kurtosis. Up-to-date information can be found at www.adni-info.org. The ADNI study was approved by the local Institutional Review Boards of all participating sites. Study subjects and, if applicable, their legal representatives, gave written informed consent at the time of enrollment for imaging data, genetic sample collection and clinical questionnaires.

ADNI3 is a continuation of the ADNI project and the imaging protocol is updated for each ADNI substudy (ADNI1, ADNI2, ADNI3). The ADNI3 MRI protocol contains multi-echo gradient echo and multi-shell diffusion acquisitions[37]. The ADNI3 database was queried for individuals with $T_1$-weighted, multi-shell diffusion MRI, and multi-echo gradient echo (2D-GRE) MRI images at the same scanning visit. From this cohort, we selected all individuals with a diagnostic status of control at the time of the visit, which included 72 control participants (48 female/24 male; mean age=73.4 years ± 7.5 years; age range: 56 years–90 years). Details regarding image acquisition parameters can be found at www.adni-info.org/methods. Imaging data were downloaded between December 2019 and January 2024.

*2.2.2 Image processing.* Susceptibility images were processed as described in the previous sections. Diffusion data were first denoised [32] then corrected for motion and eddy currents using EDDY. Next, field maps were constructed and used to correct magnetic field inhomogeneities in the diffusion images using FUGUE as previously described[38]. Mean iron and kurtosis were extracted from each grey matter ROI as described above.

### *2.3 Phantom Experiment*

An agarose phantom was designed and prepared with the goal of approximating relaxation properties in different grey matter nuclei. The concentration of agarose was chosen to mimic relaxation properties of grey matter[39] and the concentration of ferric citrate were chosen to match $R_2$* values seen in grey matter structures in the brain. A 10 mM ferric citrate stock solution was first prepared by dissolving ferric citrate powder in deionized water and mixing. Agarose powder was weighed to yield a final concentration of 2.5% in a total volume of 15 mL (0.375 g agar per phantom). Defined volumes of the 10 mM ferric citrate stock were then added to achieve final ferric citrate content of 0.03, 0.06, 0.09, and 0.12 mMol per vial. Deionized water was added as needed to bring the final volume to 15 mL. Each solution was then brought to a boil using a microwave and each solution was poured into a 15 mL vial. Following preparation, the vials were briefly sonicated to remove trapped air bubbles and rapidly cooled in an ice bath to ensure iron did not separate from the agarose mixture and settle at the bottom of the vial. A 15 mL vial containing deionized water was prepared. Finally, a 1 L liter deionized water and 2.5% agarose solution was prepared and brought to a boil in a microwave. The solution was sonicated and then poured into a 1.5 L plastic container. The vials were then embedded in a solution and images of the phantom are shown in **Figure 2**.

The phantom was scanned with a diffusion-weighted acquisition (TE/TR=70/4000 ms; voxel size=1.4 mm isotropic; 30 directions were acquired for each *b*-value [200, 400, 500, 600, 700, 1000, 1200, 1500, 1700 and 2000 s/mm$^2$] plus 5 *b*=0 images, with 10 averages per combination of direction and *b*-value). A second diffusion acquisition with identical parameters (TE/TR, spatial acquisition parameters, directions, b-values, averages) but with reverse phase encoding was acquired to correct for susceptibility distortions. Data were processed as in the human scans. Data for each shell were denoised separately in mrtrix [32], then corrected for eddy current and susceptibility distortions using eddy in FSL. MK was estimated in DIPY [35]. ROIs in were placed in each vial and a control ROI was placed in a region outside the vials and in a 20 × 20 mm$^2$ ROI at the center of the phantom across three contiguous slices. MK was measured in each ROI.

A

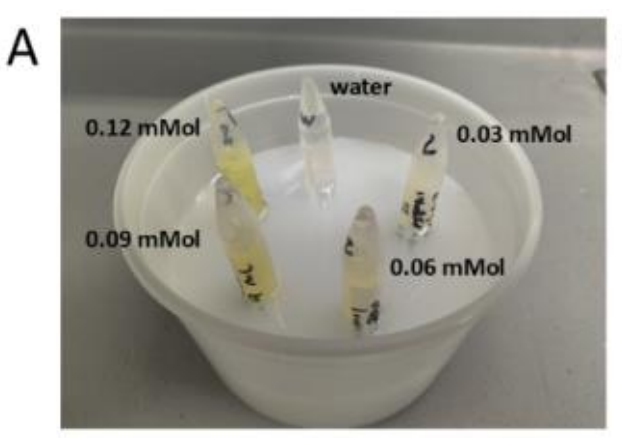

B

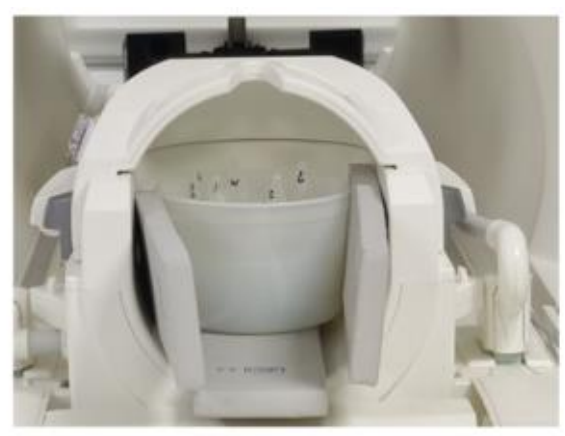

**Figure 2.** Photographs of the agarose phantom setup. (A) The container with 1 L of water mixed with 2.5% agarose. Four vials contained a mixture of water with 2.5% agarose and different concentrations of ferric citrate (0.03 mMol, 0.06 mMol, 0.09 mMol, and 0.12 mMol) and one vial contained deionized water. (B) An image of the phantom inside a 32-channel head coil.

A multiecho gradient echo pulse sequence was used for measurement of $R_2$* using the following parameters: TE1/ΔTE/TR=4/6/40 ms, FOV=192×224 mm$^2$, matrix size=192×224×96, and slice thickness=1.7 mm. $R_2$* and susceptibility maps were calculated as described above. ROIs in were placed in each vial and a control ROI was placed in a region in the center of the phantom.

### *2.4 Statistical Analysis*

All statistical analyses were performed using IBM SPSS Statistics software version 24 (IBM Corporation, Somers, NY, USA) and results are reported as mean ± standard deviation. A *P* value less than 0.05 was considered significant for all statistical tests. Age group MK and $R_2$* comparisons between the young and older cohort were made using a 2 Group (younger, older) × 9 (subcortical grey matter and white matter) Region analysis of variance (ANOVA) for MK comparison and separate 2 Group (younger, older) × 6 Region (subcortical grey matter) ANOVA for $R_2$*. Significant interactions were followed with *post hoc* between-group two-tailed *t*-tests for each region.

Within older adults, relationships between age and both MK and iron ($R_2$*) measures were assessed with separate Pearson correlations. These relationships were not assessed in younger adults due to their restricted age range.

To assess the relationship between iron and MK in subcortical grey matter, Pearson correlations between iron measures ($R_2$*) and MK in grey matter ROIs were performed separately in each age group.

In the phantom experiment, the effect of iron concentration on MK was tested with a 5 Concentration (0, 0.03, 0.06, 0.09, 0.12 mMol) one-way ANOVA. If the main effect was significant, *post hoc* comparisons between MK values in each vial were performed using respective two-tailed *t*-tests.

## 3. Results

### *3.1 Discovery Cohort*

*3.1.1. Age group differences in mean kurtosis.* Maps of MK in each age group are shown in **Figure 3**. Significant main effects of Group ($P<10^{-3}$; F=90.4) and region ($P<10^{-3}$; F=790) were seen in the 2 Group (younger, older) × 9 (subcortical grey matter and white matter) Region ANOVA. A significant interaction was observed between Group and Region ($P<10^{-3}$; F=31.65). Post hoc t-tests revealed that the older adult group had higher MK than the younger group in the putamen ($P<10^{-3}$), globus pallidus (P=0.002), caudate nucleus ($P<10^{-3}$), dentate nucleus ($P<10^{-3}$), and hippocampus ($P<10^{-3}$), but not in the thalamus (P=0.393). In white matter ROIs, the older adult group had lower MK than the younger group in the superior longitudinal fasciculus ($P<10^{-3}$), forceps major (P=0.011), and forceps minor ($P<10^{-3}$). These comparisons are shown in **Figures 4 and 5** and mean values for each group are summarized in **Table 1**.

*3.1.2 Effects of age on mean kurtosis in older adults.* In subcortical grey matter structures of the older adult group, older age was correlated with lower MK in the globus pallidus ($r$=-0.206; $P$=0.033), caudate nucleus ($r$=-0.224; $P$=0.020), thalamus ($r$=-0.264; $P$=0.006), and hippocampus ($r$=-0.269; $P$=0.005). No significant correlations were observed in the putamen ($r$=0.082; $P$=0.401) or dentate nucleus ($r$=-0.145; $P$=0.135). In white matter tracts of the older adult group, older age was significantly associated with lower MK in the superior longitudinal fasciculus ($r$=-0.442; $P<10^{-3}$), forceps major ($r$=-0.339; $P<10^{-3}$), and forceps minor ($r$=-0.370; $P<10^{-3}$).

*3.1.3 Age group differences in $R_2$*.* Maps of mean $R_2$* and susceptibility in each age group are shown in **Figure 3**. Significant main effects of Group ($P<10^{-3}$; $F$=293.350) and region ($P<10^{-3}$; $F$=620.713) were seen in the 2 Group (younger, older) × 6 (subcortical grey matter) Region ANOVA of $R_2$*. A significant interaction between Group and Region was observed ($P<10^{-3}$; $F$=36.437). Post hoc *t*-tests revealed that the older adult group had higher mean $R_2$* than the younger group in the putamen ($P<10^{-3}$), globus pallidus ($P<10^{-3}$), caudate nucleus ($P<10^{-3}$), dentate nucleus ($P<10^{-3}$), and hippocampus ($P<10^{-3}$), but not in the thalamus ($P$=0.568).

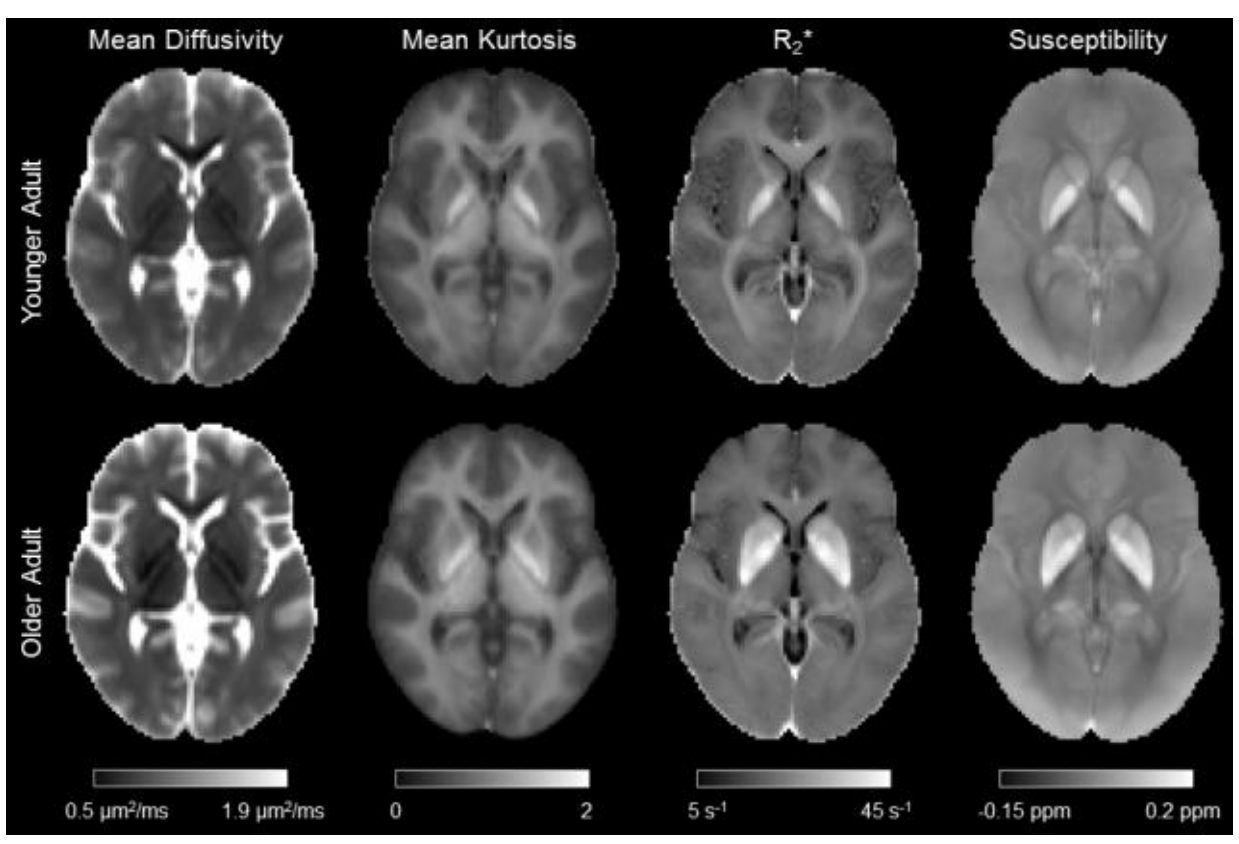

**Figure 3.** Group average images for mean diffusivity (first column) mean kurtosis (second column), $R_2$* (third column), and susceptibility (fourth column) in younger (top row) and older (bottom row) participants. These images were created by transforming each participant's mean diffusivity, mean kurtosis, $R_2$* map, or susceptibility map to Montreal Neurological Institute (MNI) common space and averaging within each group

Significant main effects of Group ($P<10^{-3}$; $F$=80.156) and region ($P<10^{-3}$; $F$=11123.864) were seen in the 2 Group (younger, older) × 6 (subcortical grey matter) Region ANOVA of susceptibility. A significant interaction between Group and Region was observed ($P<10^{-3}$; $F$=48.973). *Post hoc* t-tests revealed that the older adult group had higher mean susceptibility than the younger group in the putamen ($P<10^{-3}$), globus pallidus ($P$=0.030), caudate nucleus ($P<10^{-3}$), and dentate nucleus ($P<10^{-3}$). Lower susceptibility was observed in the thalamus ($P<10^{-3}$) and hippocampus ($P<10^{-3}$) of older adults relative to younger adults. These comparisons are shown in **Figure 4** and are summarized in **Table 2.**

*3.1.4 Effects of age on $R_2$* within older adults.* Older age was correlated with lower $R_2$* in the caudate nucleus ($r$=-0.292;

**Table 1.** Age group differences in mean kurtosis

| | Younger | Older | *t* |
|---|---|---|---|
| Putamen | 0.75 ± 0.05 | 0.95 ± 0.11 | **177** |
| Globus Pallidus | 1.23 ± 0.12 | 1.33 ± 0.23 | **10.6** |
| Caudate Nucleus | 0.65 ± 0.04 | 0.70 ± 0.59 | **44.4** |
| Dentate Nucleus | 0.99 ± 0.06 | 1.06 ± 0.10 | **22.3** |
| Thalamus | 0.88 ± 0.05 | 0.89 ± 0.06 | 0.585 |
| Hippocampus | 0.66 ± 0.03 | 0.70 ± 0.04 | **38.5** |
| Sup. Long. Fasc. | 0.99 ± 0.03 | 0.96 ± 0.05 | **-18.6** |
| Forceps Major | 0.77 ± 0.03 | 0.75 ± 0.04 | **-4.88** |
| Forceps Minor | 0.89 ± 0.04 | 0.86 ± 0.06 | **-13.8** |

*Notes.* Mean kurtosis values in subcortical grey matter regions of interest and white matter tracts. Data is presented as mean ± standard deviation. *Post hoc t*-tests were used for group comparisons of MK in each ROI from which the *t*-statistics are shown.

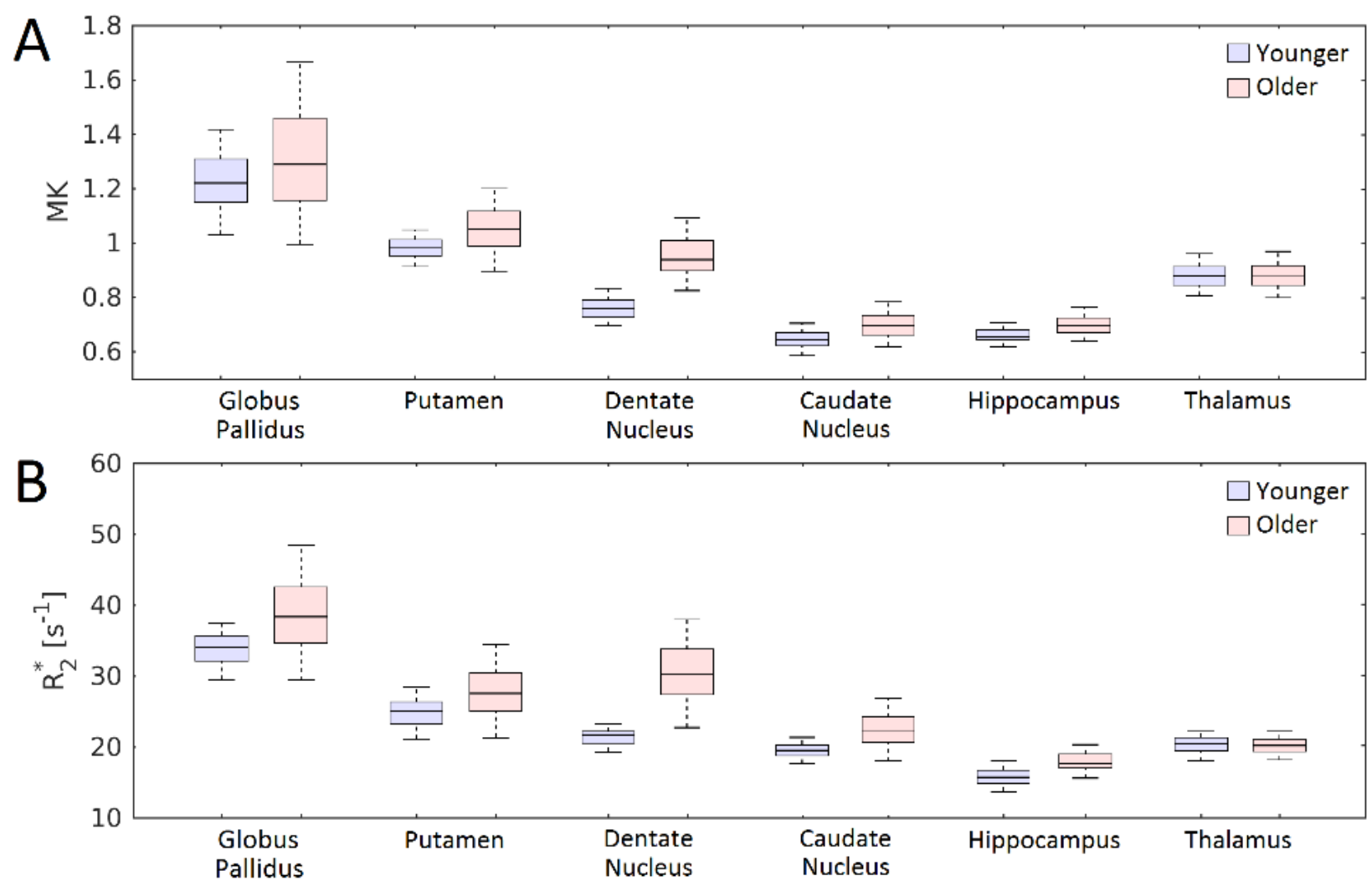

**Figure 4.** Group comparisons of mean kurtosis (MK; shown in A) and $R_2$* (shown in B) for all grey matter ROIs considered in this analysis. Significant increases in MK and $R_2$* were observed in the pallidum, putamen, dentate nucleus, caudate nucleus, and hippocampus of the older adult group relative to the younger adult group. No difference in MK or $R_2$* was seen in the thalamus.

$P$=0.001) and hippocampus ($r$=-0.194; $P$=0.044). No significant correlations between age and $R_2$* were observed in the putamen ($r$=0.064; $P$=0.517), globus pallidus ($r$=0.086; $P$=0.383), dentate nucleus ($r$=-0.118; $P$=0.230), and thalamus ($r$=-0.183; $P$=0.061). No significant correlations were seen between age and susceptibility in any grey matter ROI ($Ps$>0.061).

*3.1.5 Relationships between MK and diffusivity*. In the older adult cohort, significant negative correlations were seen between MK and MD in the superior longitudinal fasciculus ($r$=-0.834; $P<10^{-3}$), forceps major ($r$=-0.713; $P<10^{-3}$), and forceps minor ($r$=-0.712; $P<10^{-3}$) with higher MK related to lower MD. MK and MD were negatively correlated in the putamen ($r$=-0.329; $P<10^{-3}$), globus pallidus ($r$=-0.781; $P<10^{-3}$), and caudate nucleus ($r$=-0.490; $P<10^{-3}$), dentate nucleus ($r$=-0.580; $P<10^{-3}$), thalamus ($r$=-0.314; $P<10^{-3}$), and hippocampus ($r$=-0.412; $P<10^{-3}$). Similar correlations were observed in each ROI in the younger adult cohort ($Ps<10^{-3}$).

**Table 2.** Age group differences in grey matter iron metrics

| | Younger | Older | $t$ |
|---|---|---|---|
| Putamen | 21.0 ± 1.4 | 30.7 ± 5.1 | **206** |
| Globus Pallidus $R_2$* | 33.9 ± 3.3 | 39.1 ± 5.7 | **41.5** |
| Caudate Nucleus $R_2$* | 19.2 ± 1.5 | 22.4 ± 3.6 | **43.6** |
| Dentate Nucleus $R_2$* | 24.4 ± 2.7 | 28.0 ± 4.2 | **34.3** |
| Thalamus $R_2$* | 20.2 ± 1.2 | 20.1 ± 1.8 | 0.33 |
| Hippocampus $R_2$* | 15.9 ± 1.5 | 17.8 ± 1.9 | **44.0** |
| Putamen $\Delta\chi$ | 36 ± 9 | 80 ± 24 | **176.** |
| Globus Pallidus $\Delta\chi$ | 138 ± 8 | 147 ± 28 | **4.82** |
| Caudate Nucleus $\Delta\chi$ | 37 ± 9 | 48 ± 15 | **25.7** |
| Dentate Nucleus $\Delta\chi$ | 48 ± 21 | 71 ± 24 | **37.3** |
| Thalamus $\Delta\chi$ | 18 ± 6 | 4 ± 11 | **78.5** |
| Hippocampus $\Delta\chi$ | 9 ± 10 | 2 ± 11 | **16.0** |

*Notes.* $R_2$* and susceptibility, denoted $\Delta\chi$, values in subcortical grey matter regions of interest. Data is presented as mean ± standard deviation and units for $R_2$* and $\Delta\chi$ are $s^{-1}$ and ppb, respectively. *Post hoc t*-tests were used for group comparisons of $R_2$* in each ROI from which the $t$-statistics are shown.

*3.1.6. Relationships between MK and iron*. In the older adult group, MK and $R_2$* showed significant positive associations in all grey matter ROIs, including the putamen ($r$=0.702; $P<10^{-3}$), globus pallidus ($r$=0.475; $P<10^{-3}$), caudate nucleus ($r$=0.602; $P<10^{-3}$), dentate nucleus ($r$=0.667; $P<10^{-3}$), thalamus ($r$=0.202; $P$=0.038), and hippocampus ($r$=0.327; $P$=0.001). The MK-$R_2$* correlations in the older adult group remained significant after controlling for age ($Ps$<0.027). In the younger group, significant positive correlations were seen between MK and $R_2$* in the putamen ($r$=0.366; $P$=0.004), globus pallidus ($r$=0.702; $P<10^{-3}$), caudate nucleus ($r$=0.289; $P$=0.037), dentate nucleus ($r$=0.654; $P<10^{-3}$), thalamus ($r$=0.265; $P$=0.043), and hippocampus ($r$=0.283; $P$=0.030), These associations are plotted in **Figure 6**.

In the older adult group, MK and susceptibility were significantly associated in the putamen ($r$=0.606; $P<10^{-3}$), globus pallidus ($r$=0.473; $P<10^{-3}$), caudate nucleus ($r$=0.293; $P$=0.003), dentate nucleus ($r$=0.546; $P<10^{-3}$), thalamus ($r$=0.276; $P$=0.004). Controlling for age did not alter the significance of these correlations ($Ps$>0.014). No association was seen between susceptibility and MK in the hippocampus ($r$=-0.327; $P$=0.001). In the younger adult group, significant positive associations were seen between MK and susceptibility in the putamen ($r$=0.366; $P$-0.004), globus pallidus ($r$=0.827; $P<10^{-3}$), caudate nucleus ($r$=0.292; $P$=0.025), and dentate nucleus ($r$=0.567; $P<10^{-3}$) but no significant relationships between MK and susceptibility were seen in the thalamus ($r$=0.208; $P$=0.115) or hippocampus ($r$=0.071; $P$=0.594). These associations are plotted in **Figure 7.**

*3.1.7 Relationships between MK and SNR*. In the older adult group, no significant associations were seen between SNR and MK in the putamen ($r$=-0.176; $P$=0.075), globus pallidus ($r$=-0.057; $P$=0.567), caudate nucleus ($r$=-0.132; $P$=0.183), dentate nucleus ($r$=-0.181; $P$=0.067), thalamus ($r$=-0.121; $P$=0.222), or hippocampus ($r$=-0.109; $P$=0.275). Similarly, no significant associations were observed between SNR and MK in any ROI in the younger adult group ($Ps$>0.124). These associations are plotted in **Figure 8**.

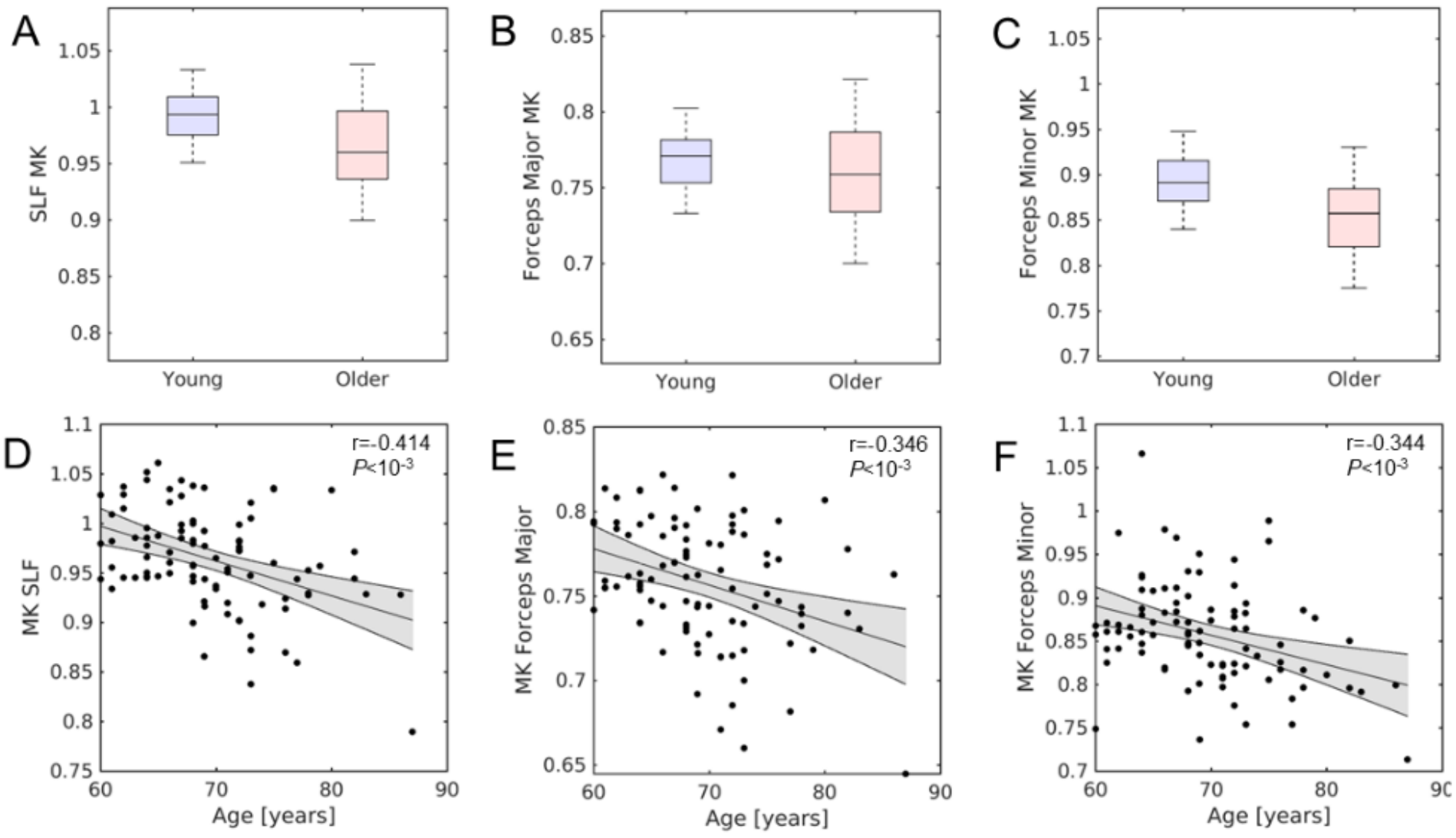

**Figure 5.** Group comparisons of mean kurtosis in the superior longitudinal fasciculus, forceps major, and forceps minor are shown in the top row (A-C). Within older adults, associations between age and MK in these white matter tracts are shown in the bottom row (D-F). Significant correlations were seen between MK and age in each white matter tract ($Ps<10^{-3}$).

### *3.2 Replication Cohort*

*3.2.1 Effects of age on MK and iron.* Significant correlations were seen between age and MK in the globus pallidus ($r$=-0.352; $P$=0.003), caudate nucleus ($r$=-0.380; $P$=0.001), dentate nucleus ($r$=-0.234; $P$=0.049), thalamus ($r$=-0.472; $P<10^{-3}$), and hippocampus ($r$=-0.318; $P$=0.008) with higher age associated with lower MK. No association was observed between MK and age in the putamen ($r$=0.135; $P$=0.266). Older age was correlated with lower $R_2$* ($r$=-0.275; $P$=0.021) and susceptibility ($r$=-0.248; $P$=0.035) in the caudate nucleus. Iron measures were not correlated with age in other grey matter ROIs ($Ps$>0.088).

*3.2.2 Relationships between MK and iron.* MK and $R_2$* were significantly positively associated in iron-rich grey matter ROIs, including the putamen ($r$=0.504; $P$=0.007), globus pallidus ($r$=0.244; $P$=0.043), caudate nucleus ($r$=0.685; $P<10^{-3}$), and dentate nucleus ($r$=0.331; $P$=0.005). The MK-$R_2$* correlations remained significant after controlling for age ($Ps$<0.042). No association between MK and $R_2$* was seen in the thalamus ($r$=0.231; $P$=0.056) or hippocampus ($r$=-0.102; $P$=0.406). Significant correlations were seen between MK and susceptibility in the putamen ($r$=0.565; $P<10^{-3}$), globus pallidus ($r$=0.309; $P$=0.010), caudate nucleus ($r$=0.480; $P<10^{-3}$), dentate nucleus ($r$=0.627; $P<10^{-3}$). No association was seen between MK and susceptibility in the thalamus ($r$=0.120; $P$=0.330) and hippocampus ($r$=0.200; $P$=0.105).

### *3.3 Phantom experiment*

The MK map for the agarose phantom with 4 vials containing different concentrations of ferric citrate and one vial of deionized water is shown in **Figure 9**. The ANOVA revealed a significant main effect of iron concentration ($F$=259.11, $P<10^{-3}$) and *post hoc* pairwise *t*-tests revealed significant differences in MK between all vials ($Ps<10^{-3}$). Mean MK in agarose without ferric citrate and in deionized water was 0.60 ± 0.01 and 0.057 ± 0.090, respectively. MK was found to increase as ferric citrate concentration increased with mean MK values of 0.65 ± 0.03, 0.90 ± 0.06, 1.13 ± 0.08, and 1.27 ± 0.09 in vials with 0.03 mMol, 0.06 mMol, 0.09 mMol, and 0.12 mMol, respectively (**Figure 9**).

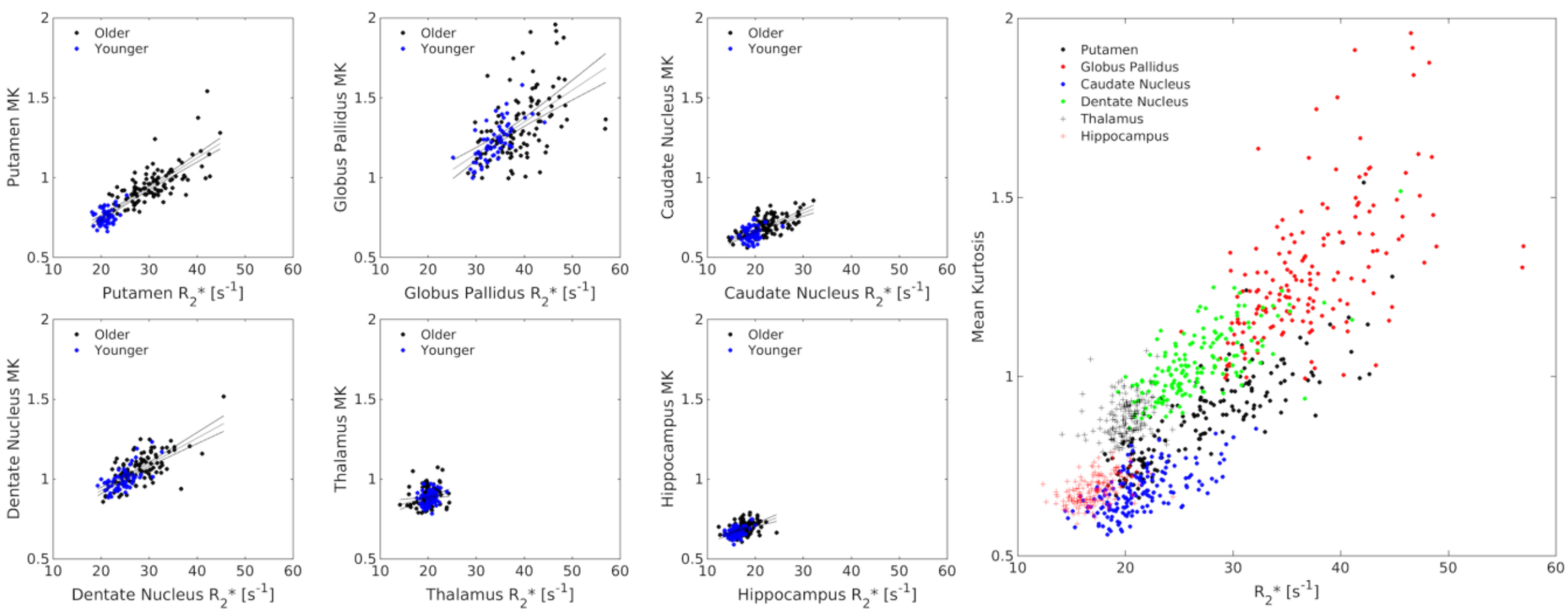

**Figure 6.** Correlations between MK and $R_2$* in the putamen (A), globus pallidus (B), caudate nucleus (C), dentate nucleus (D), thalamus (E), and hippocampus (F). The association between MK and $R_2$* in all subcortical ROIs is shown in G.

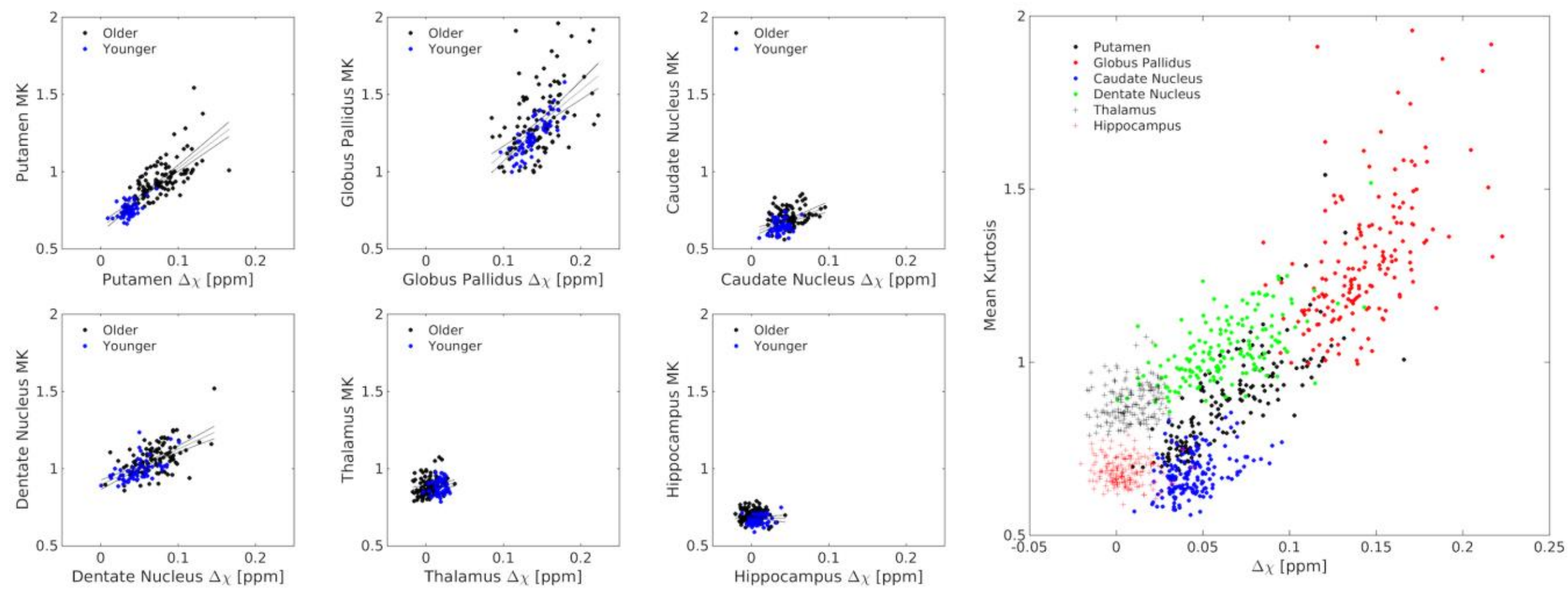

**Figure 7.** Correlations between MK and susceptibility (denoted Δχ) in the putamen (A), globus pallidus (B), caudate nucleus (C), dentate nucleus (D), thalamus (E), and hippocampus (F). The association between MK and susceptibility in all subcortical ROIs is shown in G.

## 4. Discussion

This study examined how age affects MK in subcortical grey matter structures and in white matter tracts as well as characterized the relationship between MK and iron-sensitive MRI measures ($R_2$*, magnetic susceptibility) in subcortical grey matter. In a discovery cohort, we found that age differentially affects MK in white matter and subcortical grey matter structures with lower white matter MK and higher gray matter MK in older relative to younger participants. Further, MK was significantly correlated with age in white matter tracts with lower MK associated with higher age. In subcortical iron-rich grey matter of older adults, both a discovery and replication cohort revealed that higher MK was significantly correlated to higher iron measures ($R_2$* and magnetic susceptibility). Finally, the phantom experiment found MK increases as iron concentration is increased.

Postmortem studies in humans have found age-related reductions in white matter volume and fibers in older adults[16,17]. These reductions should manifest as decreased MK[8,18]. In agreement with this, we observed older adults had, on average, lower MK in white matter tracts as compared to the younger adult group and a negative correlation between white matter tract MK and age in older adults. Prior imaging studies using DKI to examine age-related differences within white matter tracts found negative correlations between MK and age[10-14] and our correlations between MK and age in the older adult group replicate these findings.

We observed age-related increases in MK of all subcortical grey matter nuclei except the thalamus when comparing younger and older adults in the discovery cohort, but age-related decreases in MK of all regions except the putamen within the older adult groups in both the discovery (globus pallidus, caudate nucleus, thalamus, hippocampus) and ADNI (globus pallidus, caudate nucleus, thalamus, hippocampus, dentate nucleus) cohorts. Few studies have used kurtosis to examine how aging affects subcortical grey matter. Our age group comparisons are in partial agreement with at least one earlier imaging study examining age effects in MK in subcortical grey matter nuclei over the adult lifespan (25 years – 84 years), which found that age positively correlated with MK in the putamen[11]. However, whereas Gong, *et al*[11] found that age negatively correlated with MK in the caudate nucleus, globus pallidus, and thalamus in their lifespan cohort, we only observed these age-related reductions in MK within our older samples. Taken together, our findings indicate that age effects in MK vary across nuclei and may change nonlinearly across the adult lifespan. Of note, MK values in white matter were comparable to those in iron-rich subcortical grey matter nuclei. This similarity may be due the small white matter bundles that permeate striatal nuclei[40-42] and this observation is consistent with an earlier study[40].

Histological and imaging studies have found that iron accrues in subcortical grey matter nuclei throughout life[21-24]. Consistent with these results, iron-sensitive MRI metrics were found to be elevated in older adults for all grey matter nuclei except for the thalamus ROI. Iron deposition has been found to alter DTI metrics (mean diffusivity, fractional anisotropy) within iron-rich grey matter nuclei in aging and disease [25-27].

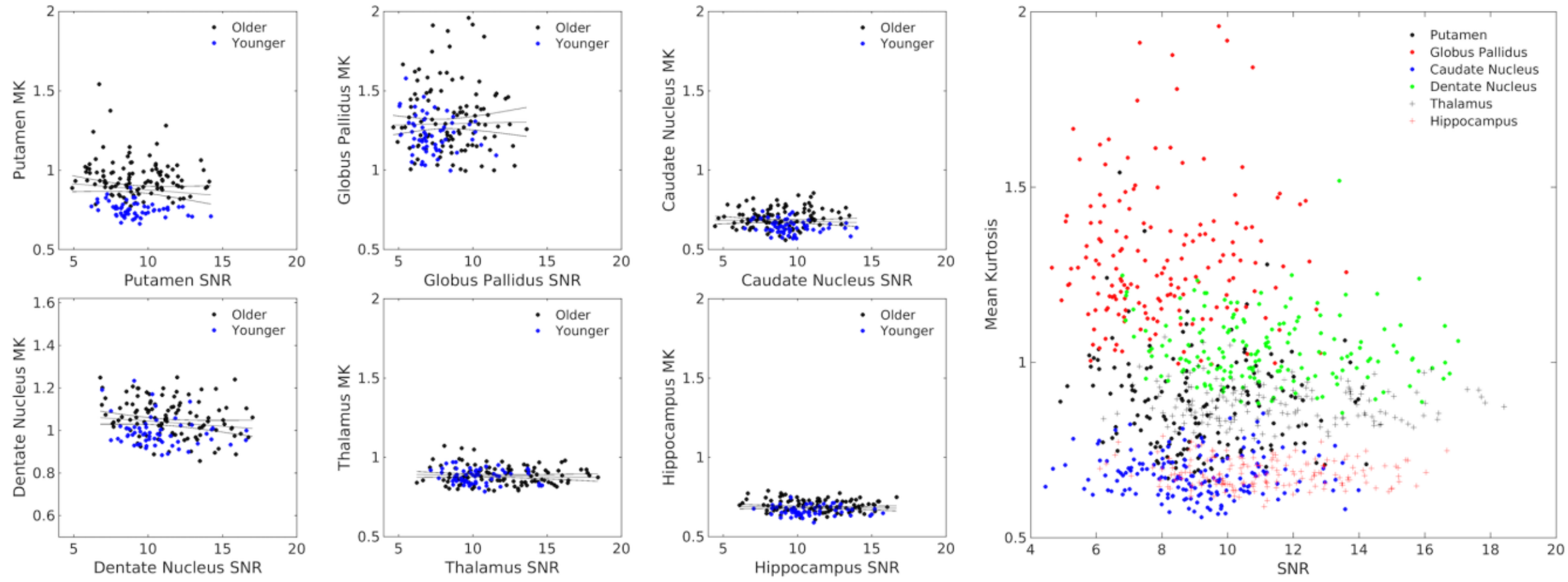

**Figure 8.** Correlations between MK and SNR from the $b$=3000 s/mm$^2$ shell in the putamen (A), globus pallidus (B), caudate nucleus (C), dentate nucleus (D), thalamus (E), and hippocampus (F). The association between MK and SNR in all subcortical ROIs is shown in G.

Consistent with these observations, we found negative correlations between iron metrics ($R_2$*, magnetic susceptibility) and mean diffusivity in all iron-rich grey matter structures (relationships shown in Figures S4 and S5). Interestingly, MK was found to be correlated with iron measures in all subcortical grey matter nuclei in the older adult discovery cohort as well as in all iron-rich grey matter nuclei in the ADNI replication cohort. In the younger adult discovery cohort, MK was correlated with iron in all grey matter structures except the thalamus.

One interpretation of the observed elevated MK values in subcortical grey matter nuclei of older adults relative to younger adults is that age-related neuronal loss and expansion of the extracellular space may increase the relative contribution of anisotropic components from these fiber bundles. These changes may lead to greater microstructural heterogeneity and increase kurtosis in grey matter nuclei of the older adult group. Alternatively, the elevated MK values in iron-rich subcortical grey matter may be due to the influence of iron. In both younger and older cohorts, structures with higher tissue $R_2$* or susceptibility values had, on average, higher MK. Consistent with this result and results showing correlations between MK and iron measures, the agarose phantom experiment found that MK increased as iron concentration increased. Iron in subcortical grey matter is not uniformly distributed in tissue but is predominantly stored intracellularly within ferritin complexes[43]. Ferritin is the primary iron storage protein in the brain and contains several thousand ferric iron atoms per molecule[44], giving rise to nanoscale paramagnetic inclusions within cells and tissue microenvironments. While ferritin does not constitute a physical diffusion barrier, its spatially heterogeneous distribution generates local magnetic field perturbations[45,46]. As water molecules diffuse through these fields, their phase accumulation becomes coupled to molecular motion, leading to deviations from a monoexponential signal decay. This coupling between diffusion and susceptibility can produce apparent non-Gaussian diffusion behavior, reflected in elevated kurtosis, even when the underlying molecular diffusion remains only weakly restricted.

The results presented here suggest iron content is related to MK. This relationship may be problematic in interpreting MK differences in neurological conditions where iron is deposited. For example, in Parkinson's disease, iron deposition occurs alongside neuronal loss in the basal ganglia [47-50] and studies have found MK is increased in the basal ganglia of populations with Parkinson's disease[19,20]. The mechanism for these increases is unknown[19] since higher kurtosis values are thought to indicate more complex tissue microstructure[8,18]. Given the relationship between $R_2$, magnetic susceptibility, and MK presented here, the increases in MK may be related to iron deposition from pathology. However, additional studies are needed to confirm that the relationship holds in pathologic populations.

This study is not without caveats. Older adults tend to have elevated iron levels as compared to younger adults[21-24] and this iron deposition will increase tissue $R_2$, thereby reducing the SNR. One limitation in the calculation of MK is SNR and low SNR may positively bias MK[51,52]. Here, we employed a denoising strategy to increase SNR[32] and mitigate the positive bias on MK. While no significant associations were observed between MK and SNR in any grey matter nuclei, we cannot rule out that noise biased MK values in the subcortical grey matter ROIs. Susceptibility values were normalized to median susceptibility in the lateral ventricles and ventricular enlargement may introduce bias. Finally, we acknowledge that the addition of iron may alter the microstructure of the agarose gel and thereby influence diffusion properties.

We examined age-related differences in MK in two older adult cohorts using gradient strengths identical to those used in the HCP Lifespan ($b$=1500 s/mm$^2$ and $b$=3000 s/mm$^2$) and ADNI3 ($b$=1000 s/mm$^2$ and $b$=2000 s/mm$^2$) studies and these results may aid in the interpretation of MK differences due to age or pathology[37,53]. We observed age-related increases in MK and iron measures ($R_2$* and susceptibility) in iron-rich grey matter structures, but age-related reductions in MK in white matter. Significant correlations were seen between MK and iron measures ($R_2$* and susceptibility) in iron-rich grey matter structures of a discovery cohort consisting of younger adults and older adults as well as in a replication cohort of older adult controls from the ADNI database. Finally, MK was found to be related to iron concentration in an agarose phantom. These findings indicate that higher MK may be related to iron content in iron-rich grey matter structures.

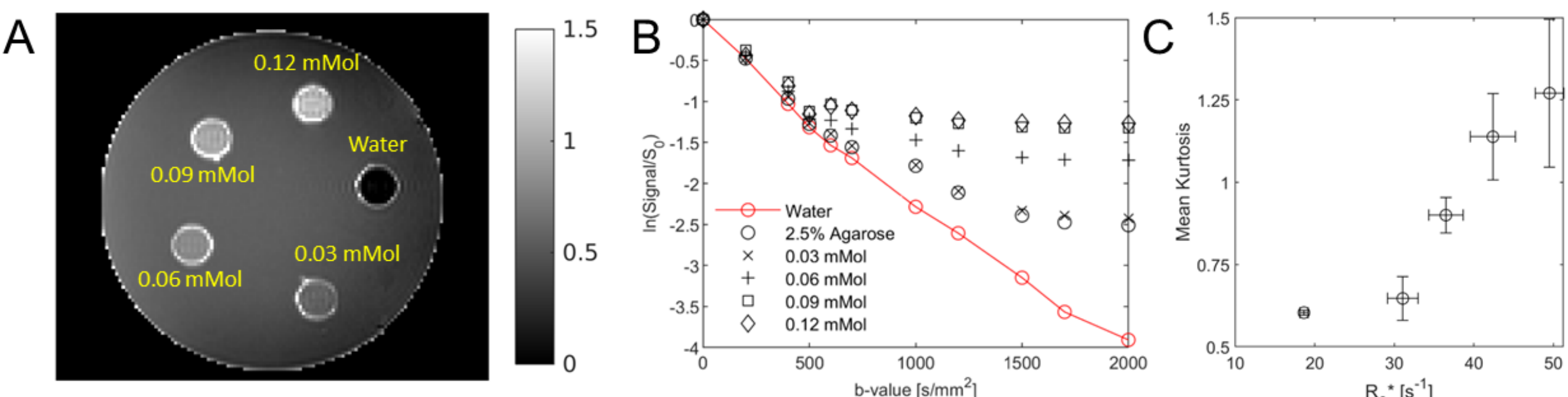

**Figure 9.** The mean kurtosis image is shown in A, a plot of the natural logarithm of the signal vs b-value for each region is shown in B, and the relationship between $R_2$* and and mean kurtosis is shown in C.

## Acknowledgements

This work was supported by R21-AG080282 and R01-AG072607 from the National Institutes of Health/ National Institute on Aging. Data collection and sharing for this project was funded by the Alzheimer's Disease Neuroimaging Initiative (ADNI) (National Institutes of Health Grant U01 AG024904) and DOD ADNI (Department of Defense award number W81XWH-12-2-0012). ADNI is funded by the National Institute on Aging, the National Institute of Biomedical Imaging and Bioengineering, and through generous contributions from the following: AbbVie, Alzheimer's Association; Alzheimer's Drug Discovery Foundation; Araclon Biotech; BioClinica, Inc.; Biogen; Bristol-Myers Squibb Company; CereSpir, Inc.; Cogstate; Eisai Inc.; Elan Pharmaceuticals, Inc.; Eli Lilly and Company; EuroImmun; F. Hoffmann-La Roche Ltd and its affiliated company Genentech, Inc.; Fujirebio; GE Healthcare; IXICO Ltd.; Janssen Alzheimer Immunotherapy Research & Development, LLC.; Johnson & Johnson Pharmaceutical Research & Development LLC.; Lumosity; Lundbeck; Merck & Co., Inc.; Meso Scale Diagnostics, LLC.; NeuroRx Research; Neurotrack Technologies; Novartis Pharmaceuticals Corporation; Pfizer Inc.; Piramal Imaging; Servier; Takeda Pharmaceutical Company; and Transition Therapeutics. The Canadian Institutes of Health Research is providing funds to support ADNI clinical sites in Canada. Private sector contributions are facilitated by the Foundation for the National Institutes of Health (www.fnih.org). The grantee organization is the Northern California Institute for Research and Education, and the study is coordinated by the Alzheimer's Therapeutic Research Institute at the University of Southern California. ADNI data are disseminated by the Laboratory for Neuro Imaging at the University of Southern California.